\documentclass[twocolumn]{article}

\usepackage{graphicx}
\usepackage{subfigure}
\usepackage{amsmath}
\usepackage{amssymb}
\usepackage{mathtools}
\usepackage{lipsum}
\usepackage{arydshln}
\usepackage{hyperref}
\usepackage{abstract}
\usepackage{color}
\usepackage{tikz}
\usepackage{cite}

\title{ \vspace{-3.5cm} \small \hfill UUITP-05/20 \\
	 \LARGE	Explore and Exploit with Heterotic Line Bundle Models}
\author{Magdalena Larfors\thanks{magdalena.larfors@physics.uu.se} , \quad 
	Robin Schneider\thanks{robin.schneider@physics.uu.se} \\
	Department of Physics and Astronomy, Uppsala University\\ SE-751 20 Uppsala, Sweden}

\begin{document}

	\twocolumn[

	\begin{@twocolumnfalse}

		\maketitle

		\begin{abstract}
		We use deep reinforcement learning to explore a class of heterotic $SU(5)$ GUT models constructed from line bundle sums over Complete Intersection Calabi Yau (CICY) manifolds.  We perform several experiments where A3C agents are trained to search for such models. These agents significantly outperform random exploration, in the most favourable settings by a factor of $1700$ when it comes to finding unique models.  Furthermore, we find evidence that the trained agents also outperform random walkers on new manifolds. We conclude that the agents detect hidden structures in the compactification data, which is partly of general nature. The experiments scale well with $h^{(1,1)}$, and may thus provide the key to model building on CICYs with large $h^{(1,1)}$.
		\end{abstract}
		\vspace{1cm}

	\end{@twocolumnfalse}
	]
	\saythanks

	\tableofcontents
	
	\section{Introduction}
	\label{introduction}

	The search for realistic four dimensional vacua coming from heterotic string theory has undergone dramatic progress in the past decades. The first handcrafted models \cite{Braun:2005ux,Bouchard:2005ag,Blumenhagen:2006ux}, have been succeeded by systematic scans over different orbifold  \cite{Lebedev:2006kn,Lebedev:2008un,Pena:2012ki} and Calabi Yau \cite{Anderson:2009mh,Anderson:2011ns,Anderson:2013xka} compactifications. There are now about 35000 models matching experimental constraints from particle physics, making this the largest set of particle physics models in heterotic string theory.\footnote{We will refer to such compactifications as 'realistic models' below.} However, these systematic approaches have also reached modern computational limits, both in the number of configurations to consider and the complexity of the involved computations. In particular, the number of configurations grows exponentially with the Hodge number $h^{(1,1)}$. In fact, $h^{(1,1)} = 6$ has more than $10^{40}$ configurations \cite{Anderson:2013xka}, and $h^{(1,1)} = 7$ has set an upper limit for systematic scans \cite{Constantin:2018xkj}. This is rather unfortunate, because at the same time it has been observed and postulated that the number of realistic configurations also grows exponentially with $h^{(1,1)}$ \cite{Constantin:2018xkj}. Thus, in order to study these unexplored fruitful regions of heterotic compactifications one has to develop different approaches to deal with the large amounts of data involved. 
	
	In the past couple of years the string theory community has started to use modern techniques from data science to deal with this issue. There are two main challenges to overcome: 
	\begin{enumerate}
		\item The large number of possible configurations to consider. The largest class of Calabi Yau manifolds is the Kreuzer-Skarke list with 473 million threefolds \cite{Kreuzer:2000xy} and $h^{(1,1)}$ ranging up to  $491$.
		The number of viable compactifications is, however, magnitudes larger. Estimates of possible configurations,  including F-theory compactifications,  range from $10^{500}$ to $10^{272.000}$ \cite{Ashok:2003gk,Halverson:2017ffz,Taylor:2017yqr,Taylor:2015xtz}.  It is obvious that this vast set cannot be studied systematically. 
		Approaches using machine learning algorithms to gain control over the vast amount of string vacua can be found in\cite{Cole:2018emh,Carifio:2017bov,Mutter:2018sra,Wang:2018rkk,Halverson:2019tkf,Halverson:2020opj}.
		\item The complexity of the mathematical problems encountered when studying compactification data. Identifying realistic models may even require the solution of NP-hard problems \cite{DENEF20071096,Cvetic:2010ky,Halverson:2018cio}. Determining Hodge numbers of manifolds and line bundles, fibration structures, number of triangulations and volumes requires computations that scale exponentially with the input.	By now there exists a large range of papers, with pioneering work \cite{Ruehle:2017mzq,He:2017aed,Krefl:2017yox}, in which such essential topological and geometrical features are being estimated by neural networks \cite{Altman:2018zlc,Klaewer:2018sfl,Brodie:2019dfx,He:2019vsj}. 
	\end{enumerate}
	
	A particularly interesting recent development is  the article \emph{Branes with Brains} \cite{Halverson:2019tkf}, where deep reinforcement learning is employed to study type IIA compactifications with intersecting D6-branes on background geometries that are $T^6/(\mathbb{Z}_2 \times \mathbb{Z}_2 \times \mathbb{Z}_{2,O})$ orientifolds. Using so-called A3C agents \cite{DBLP:journals/corr/MnihBMGLHSK16}, Ref.~\cite{Halverson:2019tkf} find models satisfying various compactification constraints, such as tadpole cancellation and a phenomenologically interesting gauge group, by,  respectively, a factor of 	200 and 20 more often than random walkers. Further, it has been demonstrated that the algorithms were able to rediscover human-derived strategies for the construction of realistic models, and also develop previously unknown strategies on their own.

	Reinforcement learning (RL) is a highly active branch of machine learning, which has, in the past few years, made major popular headlines by achieving super human performance in a variety of games. Many of these games have been previously thought to not be effectively learnable by a computer. Notably are the dominant wins against the world champion in GO, a board game with $\mathcal{O}(10^{170})$ possible board position, achieved by AlphaZero \cite{Silver1140}. AlphaZero further managed to beat other leading algorithms in chess and shogi after less than 24h of training in self-play. Hence, reinforcement learning has a proven track record of performing well in settings with large numbers, meeting one of the requirements from string theory research.
	
	Another great success was reaching competitive performance against professional players in the computer game DotA2. Here, a team of five players each controlling a hero fights against an opponent team to claim different objectives. The RL machine OpenAI-five was able to beat the current world champion in a match with slightly restricted rules \cite{OpenAI_dota}. This is remarkable, since the game itself, in contrast to GO or chess, only gives the player imperfect information. In a similar vein an agent will not know heterotic string theory nor any tools of alegbraic geometry. Further it includes random elements and has a continuous observation space with essentially infinity many allowed positions. Again, the algorithm were able to reinvent human strategies but also come up with new ones, previously unknown or deemed unfeasible by humans.
	
	Due to these strong demonstrations it certainly seems possible that RL algorithms can find novel smart ways to explore the string landscape.  Inspired by the recent success of \cite{Halverson:2019tkf}, in this report we use deep reinforcement learning and A3C agents \cite{DBLP:journals/corr/MnihBMGLHSK16} to explore and exploit a corner of the heterotic landscape. Specifically, we consider the setting of heterotic line bundle models on Complete Intersection Calabi Yau (CICY) manifolds. These heterotic configurations are, much like in the  \emph{Branes with Brains} setting, encoded by two discrete integer matrices, one describing the underlying CICY and one containing the charges of the line bundle sum.  Using the OpenAI gym package \cite{DBLP:journals/corr/BrockmanCPSSTZ16}, we implement two different environments for these agents to explore in order to identify realistic models.

	In the next section, we briefly describe the physical setting of heterotic line bundle models. We also introduce A3C agents, the overall gymCICY environment, and its reward structure. Finally, we explain the two sub-environments \emph{stacking} and \emph{flipping} developed in this study. Section \ref{sec:exp} contains the main results of our experiments, and in particular compares the number of models found by the A3Cagents against the number of models found by a random walker. In Section \ref{sec:transfer} we demonstrate that the knowledge gained on one manifold transfers to new, previously unexplored manifolds. We also employ transfer learning to further improve our results;  to our knowledge this is the first time such methods are used in the string theory community. We conclude, and comment on future prospects in Section \ref{sec:out}, and invite the interested reader to run their own experiments.  Finally, we provide a short introduction to heterotic line bundle models in Appendix \ref{ap:a}, for the benefit of readers that are not familiar with this topic. 
	
	\section{Setting}
	
	\subsection{CICYs and Line Bundles}
	\noindent
	We consider  $E_8 \times E_8$ heterotic string theory compactified on a CICY threefold \cite{Green:1986ck,Candelas:1987kf}. These are complex, K\"ahler and Ricci-flat manifolds, that are described as the vanishing locus of a set of $K=r-3$ polynomial equations in an ambient space $\mathcal{A} = \mathbb{P}^{n_0} \times ... \times \mathbb{P}^{n_r}$, where $ \mathbb{P}^{n_i}$ are projective spaces. A CICY can be described by a configuration matrix
	\begin{align}
	\mathcal{M}_{\text{compact}} =  \left[
	\begin{array}{c||ccc}
	n_0 & p^0_1 & \cdots & p^0_{K} \\
	\vdots & \vdots & \ddots & \vdots \\
	n_r & p^{r}_1 & \cdots & p^{r}_K  \\
	\end{array}
	\right]^{h^{(1,1)},h^{(2,1)}}_{\chi}.
	\end{align}
	Here, the first column $n_j \in \mathbb{Z}_{>0}$ denotes the dimension of projective spaces forming the ambient space $\mathcal{A}$, while the latter columns encode the  polynomial equations that specify the CICY.  Each $p_i^j \in \mathbb{Z}_{>0}$ gives the degree of the polynomial in the homogeneous coordinates of the projective space corresponding to that row. The Euler number $\chi$, and Hodge numbers $h^{(1,1)},h^{(2,1)}$ are topological invariants of the manifold. They are uniquely determined by the configuration matrix, but often written out for clarity.
	
	The second ingredient in heterotic GUT models is a vector bundle, whose structure group is embedded in one of the $E_8$ factors of the heterotic string, thereby specifying the gauge group and spectrum of the resulting 4D model \cite{Candelas:1985en}. We describe this in some detail in Appendix  \ref{ap:a}. Let us only state here that a vector bundle which decomposes to a sum of line bundles,
	\begin{align}
	\label{eq: V}
	V = \bigoplus_{a=1}^5 L_a = \bigoplus_{a=1}^5  \mathcal{O}(q_0^a , ..., q_r^a)
	\end{align}
	where $q_j^a \in \mathbb{Z}$, can be tuned so that it results in 4D models with $SU(5)$ GUT group \cite{Anderson:2011ns,Anderson:2012yf,Anderson:2013xka}. 
	
	The $SU(5)$ GUT group can subsequently broken to the standard model gauge group $SU(3) \times SU(2) \times U(1)$ via a Wilson line, provided that the CICY has a freely acting discrete symmetry $\Gamma$.  Hence, the physical setting we will explore in this paper is encoded in a CICY manifold $M$, a vector bundle $V$ and a  freely acting symmetry $\Gamma$. We will make use of the standard CICY list \cite{Candelas:1987kf,CICYlist} and the classification of all freely acting symmetries on CICYs descending from the ambient space \cite{Braun:2010vc} in order to provide two of these ingredients, and use reinforcement learning to generate the final ingredient, the line bundle sums $V$.
	

	\begin{figure*}[t]
		\begin{minipage}{1\linewidth}
			\centering
			\begin{tikzpicture}[shorten >=1pt,node distance=2cm,auto]
			\tikzset{
				annot/.style={
					text width=4em,
					text centered,		
				}
			}
			\tikzset{%
				every neuron/.style={
					circle,
					draw,
				},
				neuron missing/.style={
					draw=none, 
					scale=2,
					text height=0.333cm,
					execute at begin node=\color{black}$\vdots$
				},
			}
			
			\draw[dashed] (0,0) rectangle (2,8);
			\draw[dashed] (2.1,0) rectangle (8.4,8);
			\draw[dashed] (8.5,4.05) rectangle (12.2,8);
			\draw[dashed] (8.5,0) rectangle (12.2,3.95);

			\foreach \m/\l [count=\y] in {1,2,missing,4}
			\node [every neuron/.try, neuron \m/.try] (I-\m) at (1,6.5-\y) {};
			\node [annot] (Input) at (1,8.2) {Input};
			\node [annot] (IR) at (1,1) {$\mathbb{R}^{5 \cdot h^{1,1}}$};%
			
			\foreach \m/\l [count=\y] in {1,2,3,missing,4}
			\node [every neuron/.try, neuron \m/.try] (H1-\m) at (3,7-\y) {};
			\node [annot] (H1a) at (3,7) {ReLU};%
			\node [annot] (H1R) at (3,1) {$\mathbb{R}^{\text{nH}}$};%
			
			\foreach \m/\l [count=\y] in {1,2,3,missing,4}
			\node [every neuron/.try, neuron \m/.try] (H2-\m) at (4.5,7-\y) {};
			\node [annot] (H2a) at (4.5,7) {ReLU};%
			\node [annot] (H2R) at (4.5,1) {$\mathbb{R}^{\text{nH}}$};%
			\node [annot] (Hidden) at (5,8.2) {Hidden};	
			\foreach \m/\l [count=\y] in {1,2,3,missing,4}
			\node [every neuron/.try, neuron \m/.try] (H3-\m) at (6,7-\y) {};	
			\node [annot] (H3a) at (6,7) {ReLU};%
			\node [annot] (H3R) at (6,1) {$\mathbb{R}^{\text{nH}}$};
			
			\foreach \m/\l [count=\y] in {1,2,3,4,missing,5}
			\node [every neuron/.try, neuron \m/.try] (H4-\m) at (7.5,7.5-\y) {};	
			\node [annot] (H4a) at (7.5,7.5) {ReLU};%
			\node [annot] (H4R) at (7.5,1) {$\mathbb{R}^{\text{nH}+150}$};
			
			\foreach \m/\l [count=\y] in {1,2,missing,3}
			\node [every neuron/.try, neuron \m/.try] (O2-\m) at (10,4.5-\y) {};
			\node [annot] (O2R) at (11,0.5) {$\mathbb{R}^{n_L \; / \; 8 \cdot h^{1,1}}$};
			\node [annot] (O2a) at (11.5,2) {Policy};
			\node [annot] (O2l) at (11,3.5) {Softmax};
			
			\foreach \m/\l [count=\y] in {1}
			\node [every neuron/.try, neuron \m/.try] (O1-\m) at (10,7-\y) {};
			\node [annot] (O2R) at (10.5,5) {$\mathbb{R}^{1}$};
			\node [annot] (O2l) at (11.5,6) {Value};
			\node [annot] (Output) at (10,8.2) {Output};		
			
			\foreach \i in {1,2,4}
			\foreach \j in {1,...,4}
			\draw [->] (I-\i) -- (H1-\j);
			\foreach \i in {1,...,4}
			\foreach \j in {1,...,4}
			\draw [->] (H1-\i) -- (H2-\j);
			\foreach \i in {1,...,4}
			\foreach \j in {1,...,4}
			\draw [->] (H2-\i) -- (H3-\j);
			\foreach \i in {1,...,4}
			\foreach \j in {1,...,5}
			\draw [->] (H3-\i) -- (H4-\j);
			\foreach \i in {1,...,5}
			\foreach \j in {1}
			\draw [->] (H4-\i) -- (O1-\j);
			\foreach \i in {1,...,5}
			\foreach \j in {1,2,3}
			\draw [->] (H4-\i) -- (O2-\j);
			\end{tikzpicture}
		\end{minipage}
		\caption{\it The fully connected deep neural network architecture of the actor and critic, who output the policy and value function, respectively. The input states consist of the $(5,h^{(1,1)})$ integer matrices that specify the heterotic line bundle sums we want to explore. The dimensions of the policy output are $\mathbb{R}^{n_L}$ for the stacking and $\mathbb{R}^{8\cdot h^{(1,1)}}$ for the flipping environments.
		}
		\label{architecture}
	\end{figure*}
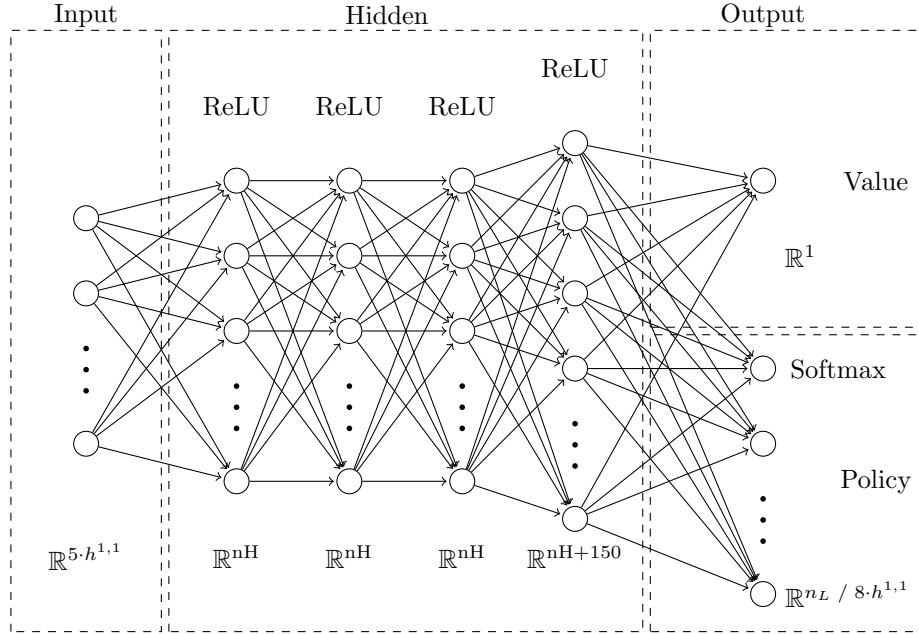

	\subsection{A3C Agents}
	
	In this section, we provide a brief introduction to reinforcement learning and the Asynchronous Advantage Actor Critic (A3C) algorithm \cite{DBLP:journals/corr/MnihBMGLHSK16}, and introduce the hyperparameters\footnote{A hyperparameter is part of the design of the experiment.} that our analysis relies on. Our discussion is far from comprehensive, and the reader is directed to Sutton and Barto's book \cite{Sutton1998} for a pedagogical and extensive introduction to the subject, and to the vast literature on reinforcement learning for more details.  We also recommend  \cite{Halverson:2019tkf}, and the insightful review by F. Ruehle \cite{Ruehle:2020jrk} for a discussion of reinforcement learning applied to problems in string theory. In particular, chapter 8 of  \cite{Ruehle:2020jrk} explains most of the concepts we need in this paper. 
	
	In short, a reinforcement learning experiment consists of an agent interacting with an environment $\mathcal{E}$, comprising of a set of states $\mathcal{S}$ and a set of actions $\mathcal{A}$. At any given time, the agent is in a state $s_t \in \mathcal{S}$, and picks some action $a_t \in \mathcal{A}$ which takes the agent to a new state $s_{t+1}$, where it receives a  scalar reward $r_t$. The agent's actions are determined by a policy $\pi$, which observes the input state $s_t$ from which it  determines  an action $a_t$. The agent is trained, {\it i.e.} updates its policy function, such that it maximizes the accumulated reward
	\[ R_t = \sum^\infty_{k=0} \gamma^k r_{t+k+1}  \, ,
	\]
	where $\gamma$ is a discount factor with $\gamma \in (0,1]$ that can be used to tune the agent's strategy towards short or long term rewards. This process continues until a terminal state $t_{end}$ is reached. 
	
	Asynchronous Advantage Actor-Critic models, proposed in 2016 \cite{DBLP:journals/corr/MnihBMGLHSK16}, provide an efficient and stable way to train reinforcement learning algorithms on a CPU. 
	Actor Critic agents consist of two estimators, usually neural networks, that are competing with each other. First, the \emph{Actor}, who is learning the policy $\pi(a_t | s_t; \theta_{\pi})$ and second, a \emph{Critic} that updates a state value function 
	\[
	V(s_t; \theta_v) =  \mathbb{E}[R_t| s = s_t] 
	\]
	giving an approximate scalar value for each state.  The policy and state value functions are determined by the weights $\theta_\pi, \theta_v$ of the involved neural networks, and the learning of the agent is encoding in updating these weights. The \emph{asynchronous} behaviour of the A3C agents arises from distributing the process over  $n_{\text{threads}}$ workers, each exploring $\mathcal{E}$ and updating their own set of parameters. Once a worker has been active for $t=t_{max}$, it will provide information that updates a global set of parameters, and then restart its training using the global parameters that encode the collective learning of all workers. The robustness of the A3C relies on these asynchronous updates of the global network from the workers.

	The updating of weights is computed via gradient descent from certain loss functions. Here A3C uses the \emph{advantage} function, which is a measure of how much better the chosen action is over others. This is estimated by taking the difference between discounted returns and the state value function. Moreover, in order to further encourage exploration one adds a measure of the entropy of the policy's actions, $H$, to the loss function. The addition  is moderated by a hyperparameter $\beta$, that thus allows to tune the agent's actions so that it favours exploration over exploitation.	Finally, in order to optimize the learning, A3C adopts the  algorithm RMSProp \cite{tieleman2012lecture} which comes with five hyperparameters: the decay factor $\alpha$, learning rate $lr$, which decays linearly, gradient clipping $gc$ and numerical stability control $\epsilon$. Please see the overview article \cite{2016arXiv160904747R} for a discussion of optimization algorithms. Ruehle's review \cite{Ruehle:2020jrk} also provides succinct explanations of different gradient descent methods and their associated hyperparamters.
	
	Our A3C agents are implemented with ChainerRL \cite{chainer_learningsys2015}\footnote{We are using version 7.2.0 for Chainer and 0.8.0 for ChainerRL.}. The layout of the neural network architectures for policy and state value function is shown in Figure \ref{architecture}. We employ a neural network where  the actor and critic only differ in the last layer: the critic has a scalar output for the value function $V(s_t;\theta_v)$, and the actor has a softmax output for the policy $\pi(a_t|s_t;\theta_\pi)$. The shared hidden layers consist of  $nH$ nodes each, and have rectified linear unit (ReLU) activation functions.  Inspired by \cite{Halverson:2019tkf}, we choose the next to last layer to have $nH+150$ nodes. Thus, the only hyperparameter controlling the architecture of the involved neural networks that we may change during experiments is $nH$, controlling the number of hidden nodes.
	
	To summarise, the reinforcement learning experiments we report on use A3C agents with several hyperparameters. Of particular importance are the discount factor $\gamma$, entropy parameter $\beta$, decay factor $\alpha$, learning rate $lr$, gradient clipping $gc$,  numerical stability control $\epsilon$ and number of hidden nodes $nH$. These hyperparamters are tuned during the design phase in order to achieve good performance of the agents; all other hyperparameters are set to their default values.

	\subsection{Environments}
	
	\label{sec:rewards}
	
	We let the agent explore two different environments,  which we, with  inspiration from  \cite{Halverson:2019tkf}, denote stacking and flipping. In both environments the observation space of the agent is the vector bundle defined in Equation \ref{eq: V}. Hence, they observe an integer matrix of shape $(5, h^{1,1})$. The agents interact with the environment by changing the line bundle charges of the vector bundle, i.e. the integer entries in this matrix.
	
	The stacking and flipping environments share a common set of machine learning hyperparameters and, in addition, several physical parameters: the upper bounds on the allowed charges ($q_{max} = 2$), the rank of the freely acting symmetry ($|\Gamma| = 2$) and the underlying CICY. 
	
	The reward structure is one of the most important properties of RL. The aim of our study is to identify line bundle sums that result in  heterotic  $SU(5)$ GUT models which at low energy reduce to the MSSM without exotics. This entails satisfying a number of physical constraints, which are summarised in the list below, and described in more detail in Appendix \ref{ap:a}.  Consequently, the agents are rewarded for finding 'good' configurations, where the 'goodness' of a configuration is determined by the number of physical constraints it satisfies. 
	
	\subsubsection*{Rewards and physical constraints}
	
	We list all physical constraints ordered according to the internal checks of the environment. Each constraint yields some reward R, if satisfied. If a constraint is not satisfied subsequent ones will not be checked. 
	\begin{enumerate}
		\item $V$ has to be an $S(U(1)^5)$ bundle, thus $c_1(V) = 0$. R$_{\text{max}}$=5. \label{c:sun}
		\item The \textit{weak} stability constraint; each line bundle has slope zero somewhere in the K\"ahler cone; $\mu(L_a)=0, a=1,...,5$. R$_{\text{max}}$=2. \label{c:weak}
		\item There are three fermionic matter generations; the index of each line bundle is in the range $-3 |\Gamma| \leq \text{index}(L_a) \leq 0$, where $|\Gamma|$ is the rank of the freely acting symmetry.  R$_{\text{max}}$= $10$. \label{c:index}
		\item There are three fermionic matter generations; the index of $V$ is determined by $\text{ind} (V) \stackrel{!}{=} -3 |\Gamma|$. R=$10^2$. \label{c:Index}
		\item The Bianchi identity and Bogomolov bounds impose $0 < c_2(V) \leq c_2 (\mathcal{M})$. R=$10^4$. \label{c:bianchi}
		\item Exclude exotic Higgs triplets; $\text{ind}(L_a \otimes L_b) < 0$. R=$10^4$.\label{c:triplet}
		\item Require at least one Higgs doublet; $h^2(\wedge^2 V) \leq 0$. 
		R=$10^6$.\label{c:doublet}
		\item There are no antigenerations, $h^2(V) \stackrel{!}{=} 0$. R=$10^7$.\label{c:fermion}
		\item $V$ needs to be slope stable somewhere in the K\"ahler cone.\footnote{We were unable to implement an efficient converging solution to solve constraint \ref{c:kaehler}, and thus omit it from our experiments. Similarly, checks of constraints related to equivariance have been omitted in our analysis. Thus, the realistic models we find are only putative in this sense.} \label{c:kaehler}
	\end{enumerate}
	The reward for the first three constraints is variable and becomes less, the further away the constraint is from being satisfied. In fact, for very bad configurations the agent gets punished and receives a negative reward. For more information about the computations and derivations of these constraints, see Appendix \ref{ap:a} and the references \cite{Anderson:2011ns,Anderson:2012yf,Anderson:2013xka}.

	\subsection{Stacking}
	
	The stacking environment comes closest to the strategy employed in \cite{Anderson:2013xka} to generate a systematic list of realistic line bundle models. Here, a list $L$ of line bundles that satisfy conditions \ref{c:weak} and \ref{c:index} is precompiled. The agent has control over the first four line bundles by picking a $L_a \in L$ and cyclically replacing an old one. The fifth line bundle is fixed by constraint \ref{c:sun}, such that $c_1(V) = 0$. An episode ends, if a standard model has been found or after maximum, $m_{\text{steps}}$, number of steps.
	
	The action space is given by, $n_L = len(L)$, and the number of possible configurations up to permutations is
	\begin{align}
	\label{eq: nconfs}
	N_{\text{conf}} = n_L^4.
	\end{align}
	The initial stack is a random sample of four line bundles $\in L$. There is no reward for the first three constraints as they become trivial in this setting. It is important to note, that after four steps the whole observation space has been changed. Thus, the agent does not need to envision a long term strategy.

	\subsection{Flipping}
	
	The flipping environment lets the agent control each charge $q_j^a$ separately. Starting from an initial random configuration that satisfies $c_1(V) = 0$ with $q_j^a \in \{-1,0,1\}$, the agent can either increase or decrease a single charge of the first four line bundles by one. This change is compensated in the fifth line bundle such that $c_1 (V)$ remains trivial. If the action increases a charge to a value larger than $q_{max}$, cyclic boundary conditions are imposed. 
	
	The action space of the flipping environment is of size $h^{1,1} \cdot 4 \cdot 2$.
	The number of possible configurations up to permutations in this setting are
	\begin{align}
	\label{eq: nconff}
	N_{\text{conf}} = (2 \cdot q_{max} +1)^{4 \cdot h^{1,1}}.
	\end{align}
	In contrast to the stacking environment only the first condition is trivially satisfied and will not yield any reward. It takes much longer to explore clearly distinct configurations, as the observation space only  changes minimally between each step. Hence, in this setting it becomes more important for the agent to develop long term strategies.
	
	\begin{figure*}[t]
		
		\centering
		\begin{minipage}{0.47\linewidth}
			\includegraphics[scale=0.5]{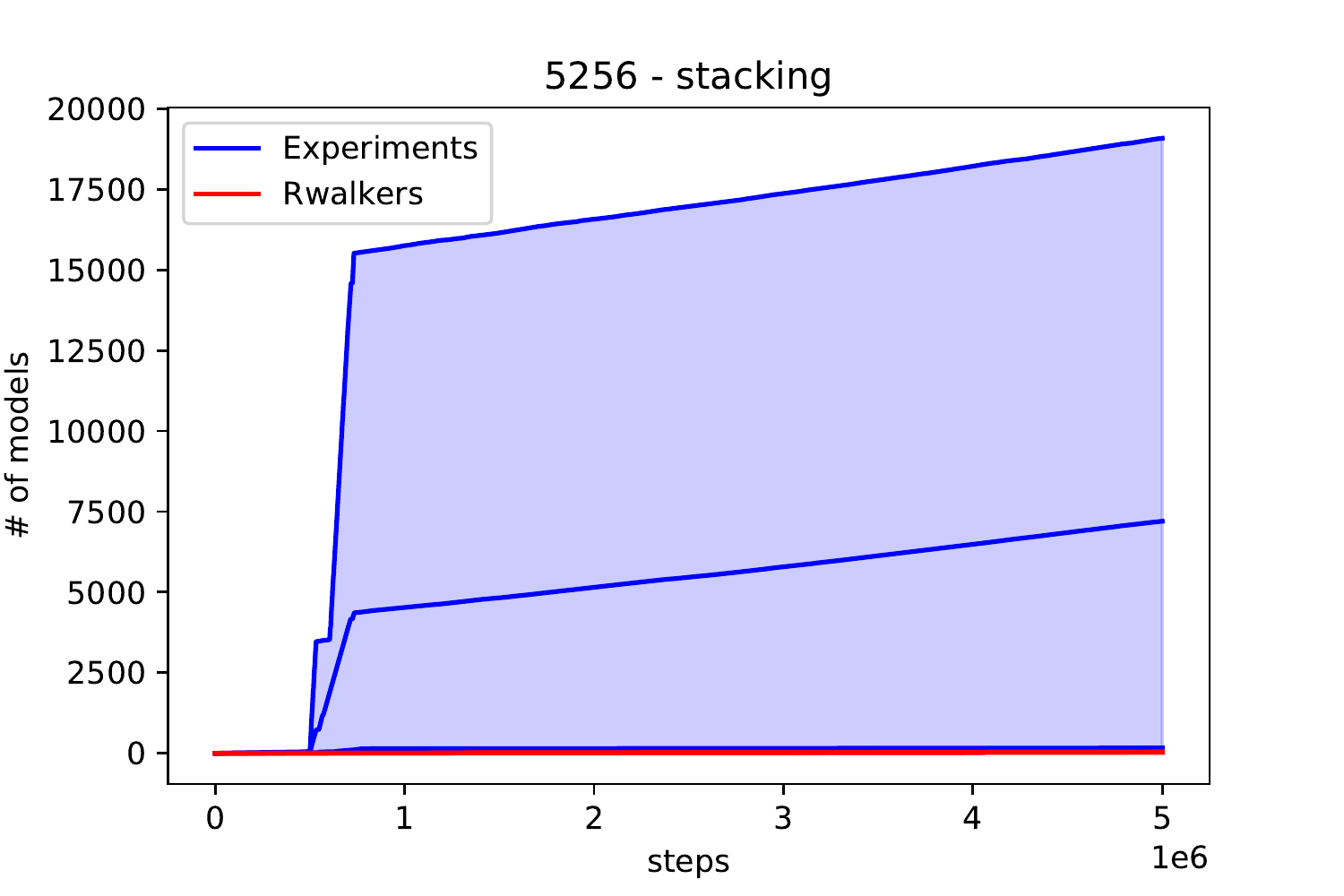}
		\end{minipage}
		\begin{minipage}{0.47\linewidth}
			\includegraphics[scale=0.5]{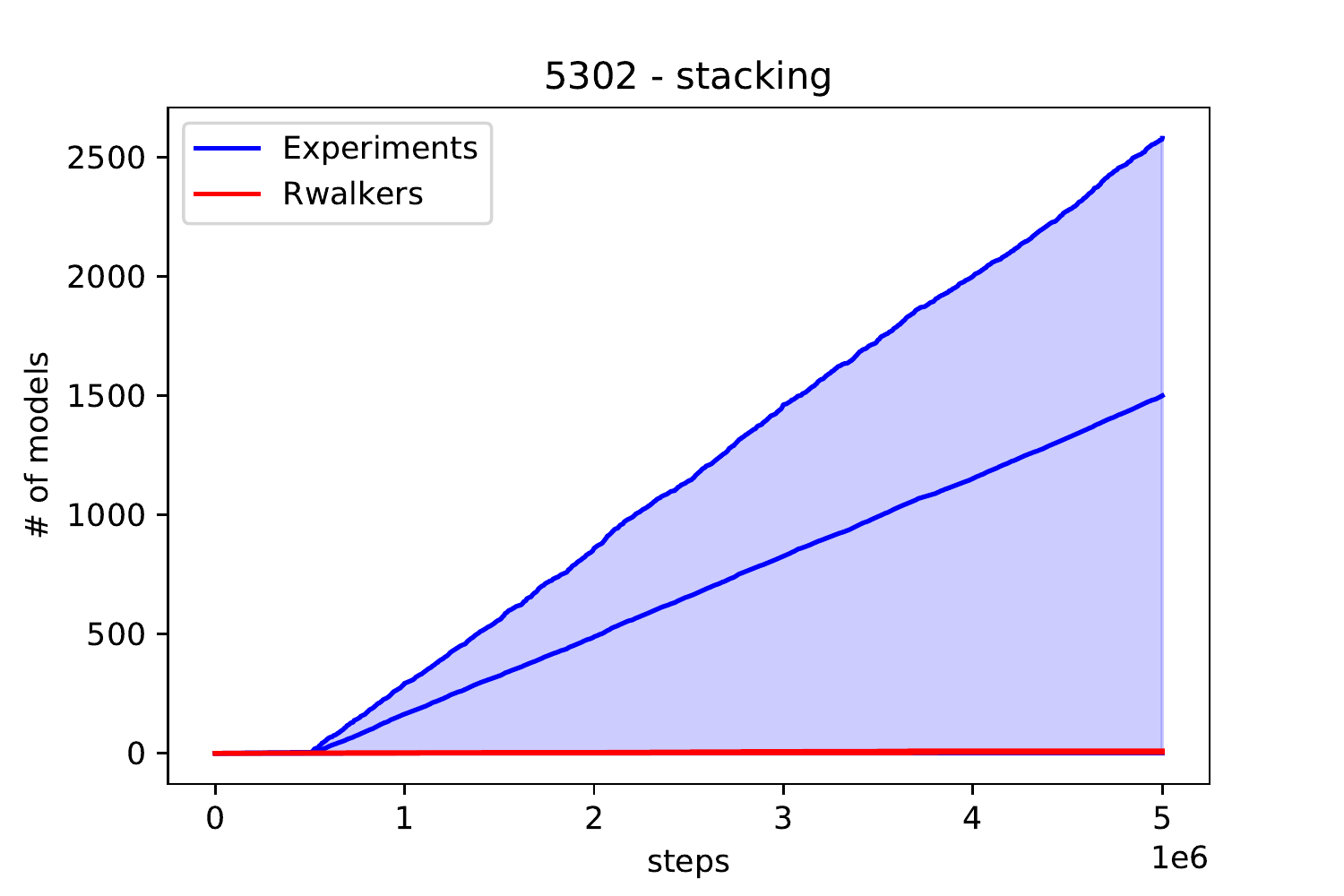}
		\end{minipage}
		
		\caption{\it Number of models found in two selected sets of stacking experiments (in blue) for the manifolds 5256 and 5302 plotted for comparison with each five random walkers (in red).}
		\label{fig: stack}
	\end{figure*}

	\section{Experiments}
	
	\label{sec:exp}
	
	The experiments were primarily performed on the two CICYs with number 5302 and 5256.  Both geometries admit freely acting symmetries of rank 2, and their configuration matrices are given respectively by
	\begin{align}
	\mathcal{M}_{5302} =  \left[
	\begin{array}{c||ccc}
	1 & 0 & 1 & 1 \\
	1 & 0 & 1 & 1 \\
	1 & 1 & 1 & 0 \\
	1 & 1 & 1 & 0 \\
	1 & 1 & 0 & 1 \\
	1 & 1 & 0 & 1
	\end{array}
	\right]^{6,30}_{-48}
	\end{align}
	and
	\begin{align}
	\mathcal{M}_{5256} =  \left[
	\begin{array}{c||cccc}
	1 & 1 & 1 & 0 & 0 \\
	1 & 2 & 0 & 0 & 0 \\
	1 & 0 & 0 & 1 & 1 \\
	1 & 0 & 0 & 1 & 1 \\
	3 & 1 & 1 & 0 & 1
	\end{array}
	\right]^{5,29}_{-48}.
	\end{align}
	According to \cite{Anderson:2013xka} these CICYs  are both fruitful patches for heterotic line bundle models.  
	
	The studies were performed on the NSC cluster Tetralith using two Intel® Xeon® Gold 6130 CPUs with each 16 cores totaling 32 workers and 96 GiB available memory. Experiments were terminated after a predefined number of total steps, $t_{\text{steps}}$. Our average computation time was 50 minutes for a single stacking experiment and about 3 hours and 30 minutes for a flipping experiment.
	
	All experiments share a set of common values for some of the hyperparameters: $q_{max} = 2$, $\alpha = 0.99$, weight decay $=0$,  $t_{max} = 5$, $|\Gamma| = 2$, $gc = 20$ and $\epsilon = 0.0001$. Other hyperparameters, listed in  Table \ref{tab:hyper}, are tuned in order to improve performance, through a set of preparatory experiments. We then picked the most promising ones and subsequently repeated the experiment with four different seeds. Hyperparameters that are not explicitly mentioned are set to default values.
	\begin{table}[h!]
		\caption{\it Remaining hyperparameter values of the various experiments.}
		\begin{center}
			\begin{tabular}{|c|cc|cc|}
				\hline
				& stack & &flip& \\
				parameter & 5256 & 5302 & 5256 & 5302\\
				\hline 
				$t_{\text{steps}}$ & $5 \cdot 10^6$ &$5 \cdot 10^6$ &$50 \cdot 10^6$ &$50 \cdot 10^6$ \\
				$m_{\text{steps}}$ & 30 & 30 & 200 & 300\\
				$lr$ & $5 \cdot 10^{-4}$ & $5\cdot 10^{-4}$ & $10^{-4}$
				& $10^{-4}$\\
				$nH$ & 100 & 100 & 75 & 100\\
				$\beta$ & 1 & 0.1 & 1 & 1\\
				$\gamma$ & 0.7 & 0.7 & 0.95 & 0.95\\
				\hline
			\end{tabular}
		\end{center}
		\label{tab:hyper}
	\end{table}
	
	Due to the asynchronous updating of the weights defining the agents, it is not possible to exactly reproduce the results of a previous experiment even when one has exact knowledge of all involved seeds. In fact the results can differ greatly from experiment to experiment. This is a well-studied subtlety of asynchronous RL algorithms, see 
	\cite{henderson2017deep,islam2017reproducibility} and the interesting blog post by Alex Irpan \cite{rlblogpost}. Due to these complications we followed the practice suggested in \cite{henderson2017deep,islam2017reproducibility}, and report results for five experiments of different seeds for each choice of hyperparameters. 
	
	In the next sections we present the results of our experiments by evaluation of how many realistic models have been found after $t$ time steps. This is different from the standard approach of plotting the accumulated reward in an episode. The motivation for this choice is that the number of models is the only quantity we are interested in from a physics perspective. We note though, that this can lead to disappointing results in some experiments, where the agents get stuck taking actions that maximizes the reward, but does not find any models.

	\begin{table}[t]
		\begin{center}
			\begin{tabular}{|c|c|c|}
				\hline
				manifold & experiment & \# of models \\
				\hline
				5302&flip-best & 16831 (14137)  \\
				5302&flip-mean & 12866.6 (11165.4)  \\%
				5302&rw-mean & 8.4 (8.4)  \\
				\hline 
				5302&stack-best & 2579 (272)  \\
				5302&stack-mean &  1499.4 (151)  \\
				5302&rw-mean & 6.6 (6.6)  \\
				\hline
				5256&flip-best & 8122 (2125)  \\
				5256&flip-mean & 5160 (1722.4) \\%
				5256&rw-mean & 6.4 (6.4)\\%
				\hline
				5256&stack-best & 19087 (165)  \\
				5256&stack-mean & 7201.6 (107)  \\ 
				5256&rw-mean & 37 (37)  \\ 
				\hline
			\end{tabular}
		\end{center}
		\caption{\it Number of putative models for mean, random walker and best performing stacking and flipping experiments on the manifolds 5256 and 5302. Brackets denote the number of models after removing duplicates and permutations of the line bundles.}
		\label{t: nmodels}
	\end{table}
	
	\begin{figure*}[ht]
		
		\centering
		\begin{minipage}{0.45\linewidth}
			\includegraphics[scale=0.5]{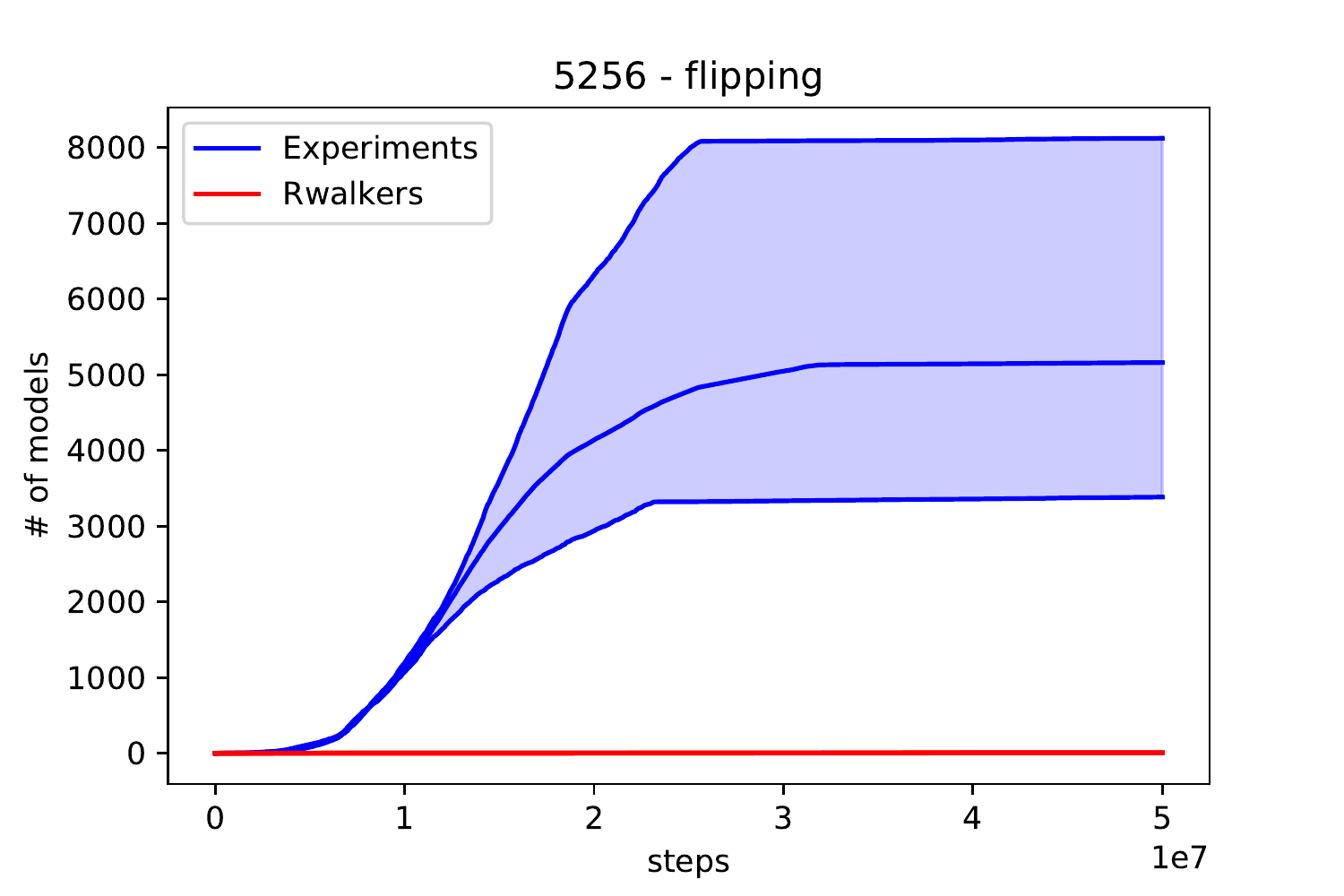}
		\end{minipage}
		\begin{minipage}{0.45\linewidth}
			\includegraphics[scale=0.5]{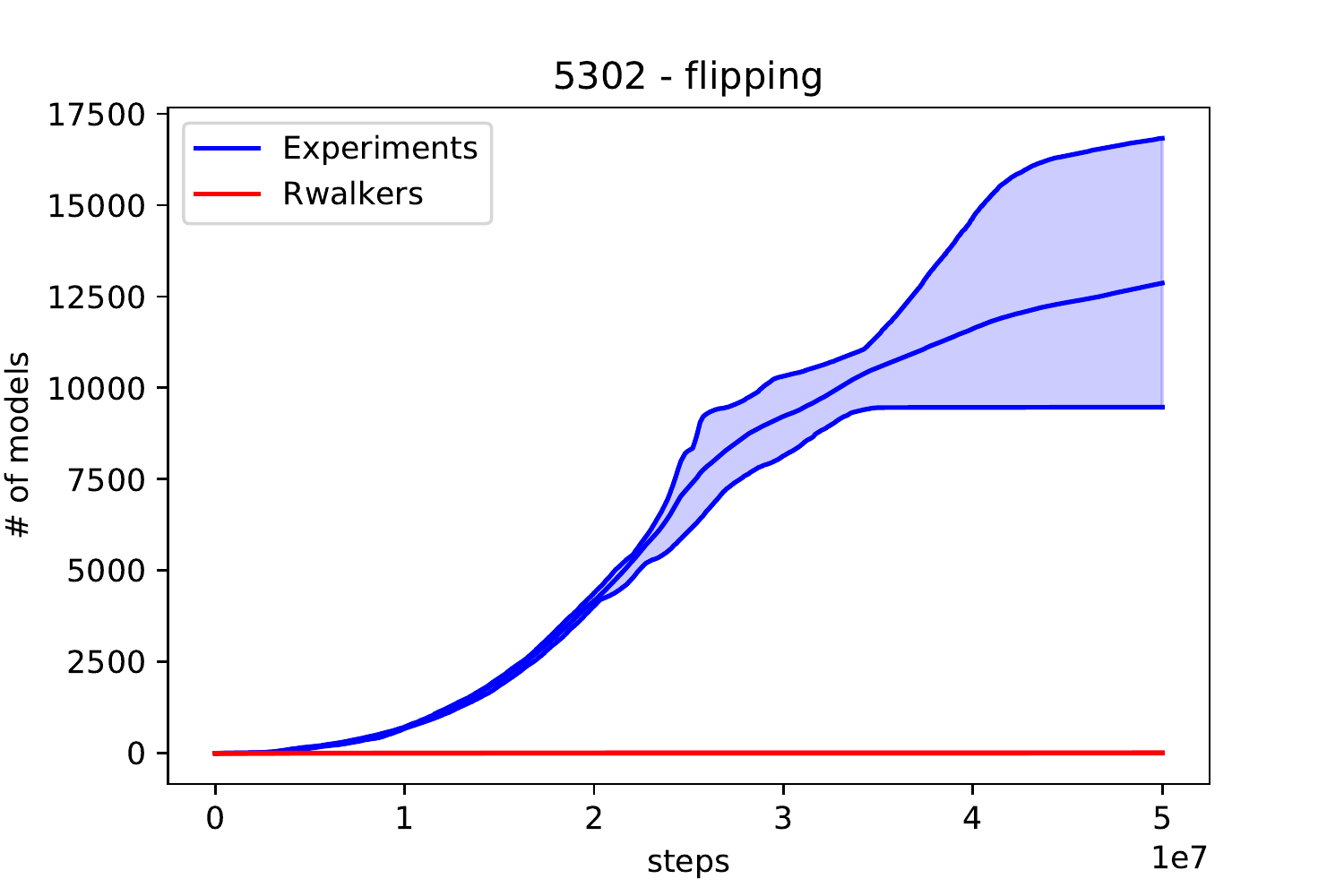}
		\end{minipage}
		
		\caption{\it Number of models found in two selected sets of flipping experiments (in blue) for the manifolds 5256 and 5302 plotted for comparison with each five random walkers (in red).}
		\label{fig: flip}
	\end{figure*}
	
	\subsection{Stacking}
	
	In Figure \ref{fig: stack} we plot the number of models in two different sets of stacking experiments on 5256 and 5302. 	In blue are the results of five experiments with the same hyperparameters, where the top line indicates the best performing experiment, the middle one the mean and the bottom one is the worst performing experiment. For comparison we added results of five differently seeded random walkers in red. 
	Table \ref{t: nmodels} shows absolute numbers for the mean and best performing experiments.
	
	We observe that the two blue regions cover the entire range down to the random walkers. In fact the worst performing experiment on 5256 finds only three times more models than a random walker. For 5302 we have the worst experiment finding a single model and thus being outclassed by simple random exploration. On the other hand the best performing experiments find respectively, 19087 and 2579 models, which is a factor of 500 and 400 
	better than a random exploration. We recall, that this larger variance in performance for the same hyperparameters makes it important to report results for a set of experiments. Here, respectively, 1 in 5 and 2 in 5 of the same stacking experiment over 5256 and 5302 have effectively been failures.
	
	Studying the preferred actions and found models one finds that the agents, after some initial exploration, get stuck repeatedly picking the same action (line bundle). This line bundle has, presumably, been identified as a member of configurations yielding a large reward. Unfortunately, it does not follow that this particular line bundle is also good for model building, as can be observed by the bottom border of the blue region.
	In some cases, on the other hand, the line bundle is in fact a member of various realistic models. This accounts for the top ranges of the blue region. When such a line bundle has been identified the number of found models subsequently increases linearly until it reaches the termination condition.
	As a consequence of this behaviour, we opted to run stacking experiments only for $5 \cdot 10^6$ steps.
	
	Furthermore, this tendency to get stuck in a local minimum also explains the low number of unique models, which are found after removing all duplicates and accounting for permutations in the line bundle sum, {\it cf.}~Table \ref{t: nmodels}. Summarizing the results of Table \ref{t: nmodels} we find that in average the agents outperform random walkers by a factor of respectively $3$ and $20$ when it comes to finding unique models, but that the absolute number of unique models is not high.

	\subsection{Flipping}

	Figure \ref{fig: flip} shows the number of found models {\it vs.} the number of steps for two sets of flipping experiments on the manifolds 5256 and 5302 running for 50$\cdot 10^6$ steps. The corresponding numbers for best performing experiments, mean and random walkers are again found in Table \ref{t: nmodels}.
	In contrast to the stacking experiments, we observe that the agents take longer time to start finding realistic models, but that they consistently outperform random walkers in the long run.

	It is noteworthy that the two flipping plots have dissimilar shapes and are also different from the stacking plots.
	In both cases the experiments take about $5 \cdot 10^6$ steps, {\it i.e.} the maximal number of steps for stacking experiments, before becoming significantly better than a random walker. They then show an exponential growth in finding models. 
	For the manifold 5256 the growth becomes linear after passing the $15 \cdot 10^6$ step mark and starts to flatten at $22 \cdot 10^6$ steps. 
	
	The spread in performance is lower for the 5302 experiments. Here, the exponential growth lasts approximately to the $22 \cdot 10^6$ mark, with all experiments showing equal performance. Afterwards, the number of found models becomes more linear with the top performer going through various stages of different slope. The bottom experiment becomes a flat line similar to the 5256 experiments.
	
	The observed flat lines late in these experiments is unexpected.   Since the performance of the flipping experiments is overall very satisfactory, we leave systematic experiments which test the flatline behaviour for future work.

	According to Equation \ref{eq: nconff} there are respectively $\mathcal{O}(10^{14})$ and $\mathcal{O}(10^{16})$ possible configurations in the flipping environment. More interestingly, according to  \ref{t: nmodels}  respectively 2125 and 14137 unique models have been found in the best performing experiment for 5256 and 5302. On the other hand going through the database associated to \cite{Anderson:2013xka} we find that there are 457 and 2750 models within our range of charges satisfying all nine constraints. The discrepancy can, in part, be explained by the lack of constraint \ref{c:kaehler} for our models. We investigated all unique putative models explicitly and found that respectively 285 and 3780 of the top performing experiments satisfy constraint \ref{c:kaehler}. 
	
	The remaining discrepancy between 3780 found models and 2750 expected ones is explained by the fact that the authors of \cite{Anderson:2013xka} removed configurations with respect to permutation symmetries of the ambient space, which we have not done. 
	In general we find that in average 16 \% and 27 \% of the putative models found during our experiments satisfy \ref{c:kaehler}. Summarizing the results of Table \ref{t: nmodels} we find that random walkers are outperformed by factors of $300$ and $1700$.

	\section{Transfer of Knowledge}
	
	\label{sec:transfer}

	In the previous section we showed that A3C agents are able to consistently and significantly outperform random walkers in finding realistic string vacua in the flipping environments. While this by itself is a nice result, significant time has been spent finding good values for the hyperparameters. Thus, the upside of using reinforcement learning only really kicks in after considerable initial time and computation investment. This leads to the questions whether the knowledge obtained from one manifold can be used to identify realistic models on a new manifold. 
	
	\begin{figure*}[t]
		
		\centering
		\begin{minipage}{0.45\linewidth}
			\includegraphics[scale=0.5]{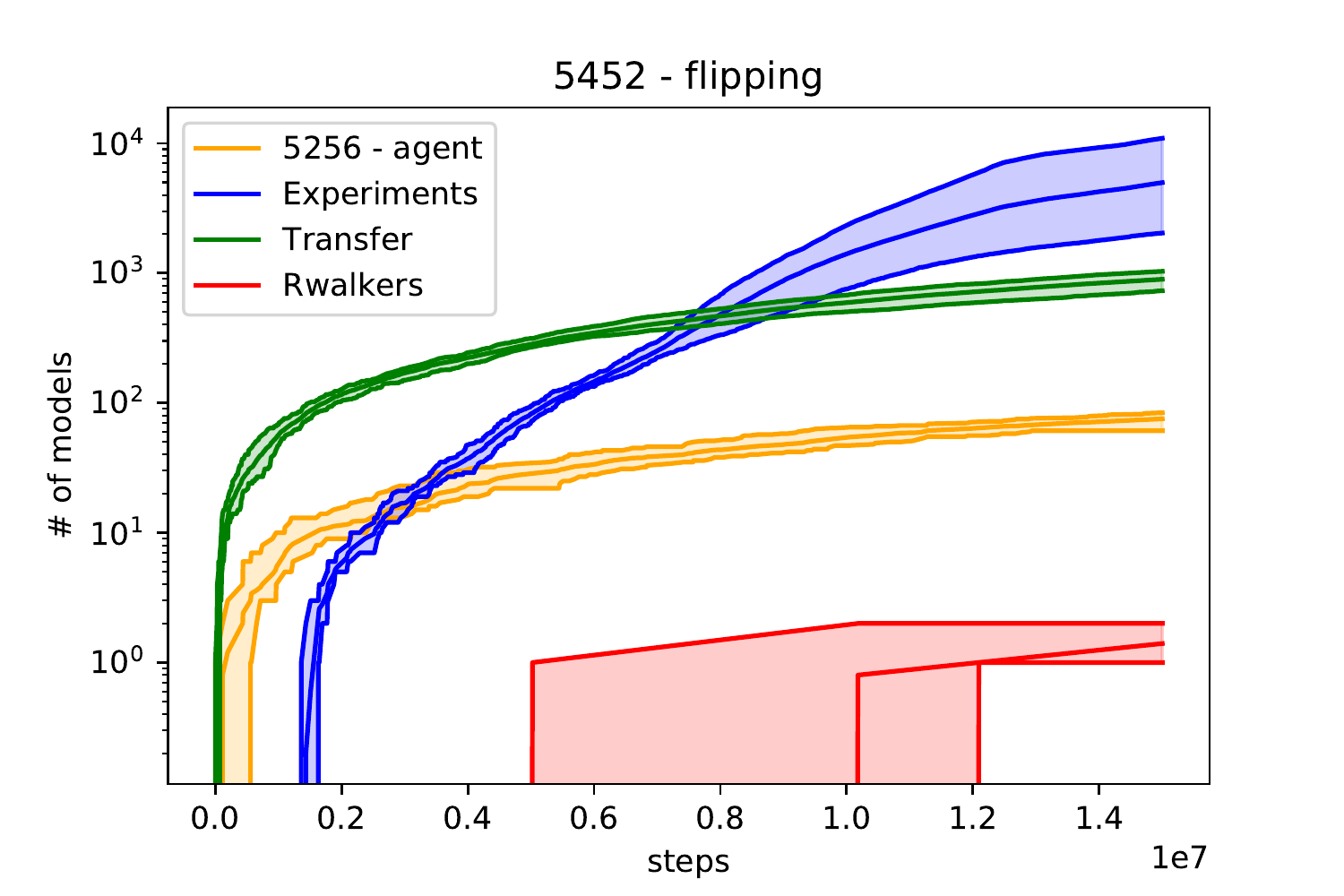}
		\end{minipage}
		\begin{minipage}{0.45\linewidth}
			\includegraphics[scale=0.5]{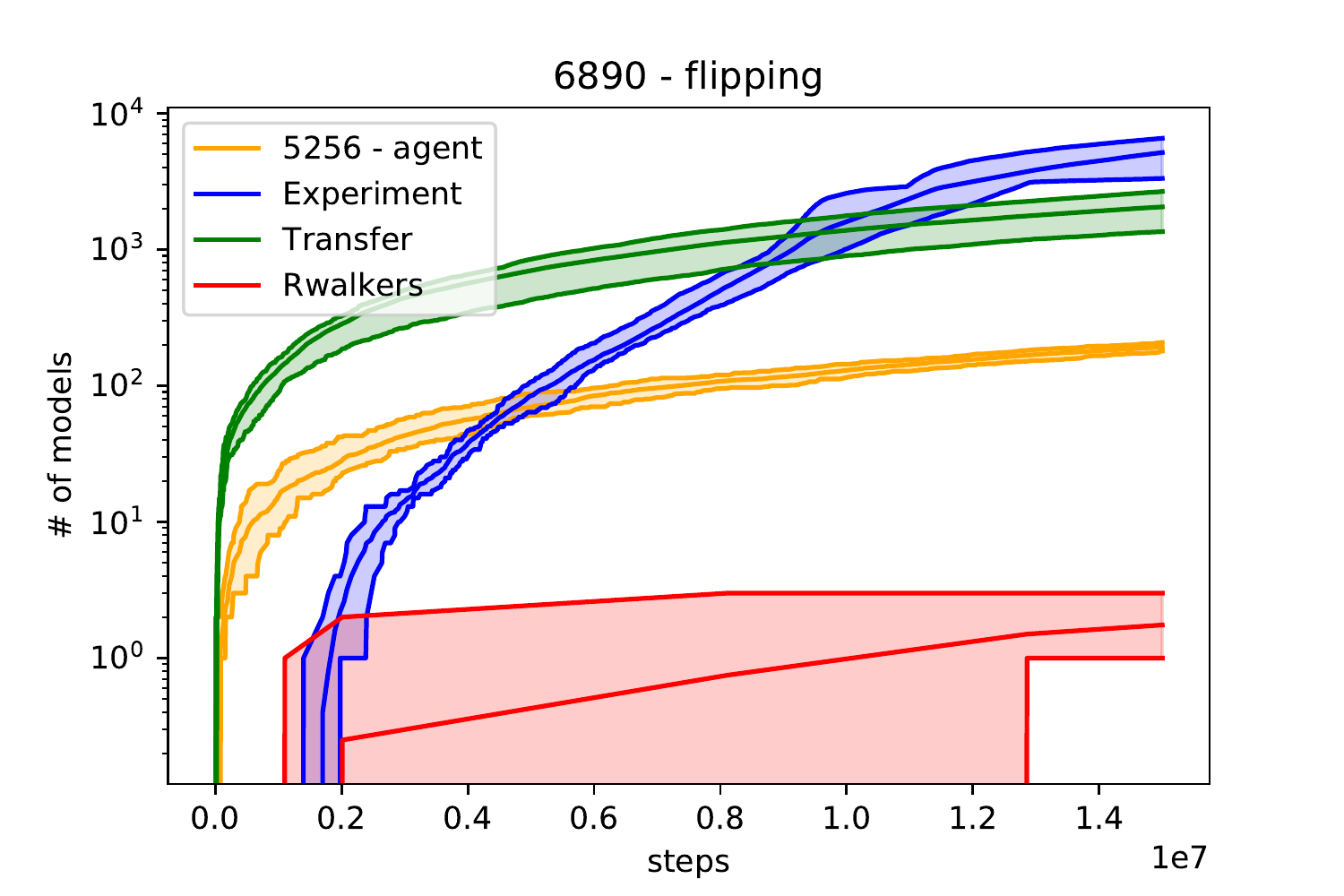}
		\end{minipage}
		
		\caption{\it Number of models found for selected sets of flipping experiments (in blue), random walkers (in red), pretrained agents (in yellow), and transfer agents (in green) on the manifolds 5452 and 6890. Note the logarithmic scale on the y-axis.}
		\label{fig: transfer}
	\end{figure*}
	
	In this section we will initiate an investigation in how far our previously trained agents generalize to a new setting, i.e. we will take the best performing agent of the 5256 flip experiment and see how it performs on a different manifold. We note that according to Figure \ref{fig: flip} the final agent of the best performing 5256 is not very good at finding realistic models. According to the plot, we had the best performing agent somewhere in the range of $12 \cdot 10^6$ - $22 \cdot 10^6$ time steps. In order to find and save the best agent, we had the 5256 flipping experiments perform ten evaluation runs every 50000 steps. We then save the agent with the highest mean reward during those runs. In our experiments this occurred at $21 \cdot 10^6$ steps.
	
	For the experiments in this section we will study the manifolds with numbers 5452 and 6890 in the CICY list, with configuration matrices
	\begin{align}
	\mathcal{M}_{5452} =  \left[
	\begin{array}{c||cccc}
	1 & 1 & 1 & 0 & 0 \\
	1 & 0 & 0 & 1 & 1 \\
	1 & 2 & 0 & 0 & 0 \\
	1 & 0 & 0 & 2 & 0 \\
	3 & 1 & 1 & 1 & 1
	\end{array}
	\right]^{5,29}_{-48}
	\end{align}
	and
	\begin{align}
	\mathcal{M}_{6890} =  \left[
	\begin{array}{c||ccccc}
	1 & 1 & 1 & 0 & 0 & 0\\
	1 & 0 & 0 & 1 & 1 & 0\\
	1 & 0 & 0 & 0 & 0 & 2\\
	1 & 0 & 0 & 2 & 0 & 0\\
	4 & 1 & 1 & 1 & 1 & 1
	\end{array}
	\right]^{5,37}_{-64} \; .
	\end{align}
	We note that the manifold 5452 can be reached from 5256 via so-called ineffective splitting \cite{Candelas:1987kf}, whereas 6890 is clearly topologically distinct.
	
	Additionally, we will employ a commonly used procedure of machine learning called transfer learning \cite{Pan2010ASO}. For a discussion of transfer learning in reinforcement learning we refer to \cite{Lazaric2012}. In essence, in transfer learning one takes a pretrained neural network, and fixes the initial layers in the hope that they capture some general properties of the input space. The weights of latter layers in the network remain free, and are adjusted in the training of the agent. Here, one has the option to start with some newly initialized weights in the last layer or the ones found from the pretrained network. In our experiments we opted for the latter choice.
	
	Figure \ref{fig: transfer} shows the results for experiments conducted in the flipping environment on the manifolds 5452 and 6890, run for 15$\cdot 10^6$ steps.  As in the previous section we have in red random walkers and in blue experiments starting from some randomly initialized agents. In addition there is also a yellow region indicating the performance of the pretrained agent on 5256 in evaluation mode for five different seeds and in green the results for a 'transfer' agent. The hyperparameters for the regular A3Cagents in blue have been chosen to be equivalent to those used in the 5256 flipping experiment, see Table \ref{tab:hyper}, and some adjusted hyperparameters for the transfer agents are presented in Table \ref{tab:hypertransfer}. In the transfer experiments, the first two layers of the agents are kept fixed, so that only the last three are updated. For these runs, we opted to pick a low learning in order to only minorly adjust the weights to the new manifolds.
	\begin{table}[h!]
		\caption{\it Hyperparameters of the transfer experiments.}
		\begin{center}
			\begin{tabular}{|c|cc|}
				\hline
				parameter & 5452 & 6890 \\
				\hline 
				$lr$ & $10^{-5}$ & $10^{-5}$\\
				training steps & 3$\cdot 10^6$ & 3$\cdot 10^6$ \\
				eval steps & 12$\cdot 10^6$ & 12$\cdot 10^6$\\
				\hline
			\end{tabular}
		\end{center}
		\label{tab:hypertransfer}
	\end{table}
	
	The blue region is reminiscent of the flipping experiments on 5256, where the agents started to significantly outperform random walkers at 5$\cdot 10^6$ steps with a subsequent flatlining of growth. Due to their fixed weights the yellow pretrained agents find as expected new models in a linear fashion. According to Table \ref{t: tnmodels} they outperform random walkers by a factor of 50 when it comes to finding unique models. While the absolute numbers are not impressive, Figure \ref{fig: transfer} indicates that the pretrained A3C agent developed a strategy in walking through the different vacuum configurations that is partly independent of the underlying manifold. This result is somewhat expected for the manifold 5452, as it is related to 5256 via ineffective splitting, but it is rather remarkable for the topological clearly distinct manifold 6890. 	
	
		\begin{table}[t]
		\begin{center}
			\begin{tabular}{|c|c|c|}
				\hline
				manifold & experiment & \# of models \\
				\hline
				5452& 5256 - trained & 75.4 (50.6)  \\
				5452& regular & 4975.2 (1018)  \\
				5452& transfer & 895.8 (193.8) \\
				5452& random w. & 1.4 (1.4)  \\
				\hline
				6890& 5256 - trained & 193.2 (54.4)  \\
				6890& regular &  5160.4 (963.4) \\
				6890& transfer & 2058.4 (141.4)  \\
				6890& random w. & 1.4 (1.4)  \\
				\hline
			\end{tabular}
		\end{center}
		\caption{\it Number of putative models for pretrained, regular, transfer agents and random walkers in the flipping environment on manifolds 5452 and 6890. Brackets denote the number of models after removing duplicates and permutations of the line bundles.}
		\label{t: tnmodels}
	\end{table}
	
	Furthermore, it shows that the 'transfer' agents, in green, improve the results of the pretrained agents after only a brief initial time of training. In fact, studying the first three million steps, we find that the transfer agents start finding models from the beginning of their training period. In contrast, the regular experiments, which find their first models only after exploring for $2 \cdot 10^6$ steps. The variance of the green cone is, as expected, larger than the yellow cone, because of the five distinct neural networks. The worst performing 'transfer' experiment performs on the same order as the pretrained ones. For a longer than $3 \cdot 10^6$ steps training period, we would expect the transfer agents to continue increasing their performance. However, in this section we are particularly interested in experiments with a low number of training steps. Table \ref{t: tnmodels} shows that the 'transfer' agents improve the results in finding unique models of the pretrained agents by about $283\%$ (5452) and $160\%$ (6890). The transfer agents appear to be slightly overfitting to known good configurations, resulting in a lower percentage of unique models compared to the regular flipping experiments. A higher value for the regularization parameter $\beta$ could counteract this behaviour.
	
	Figure \ref{fig: transfer} shows that the results of the 'transfer' agents are overshadowed by the great performance of the regular experiments. After roughly 8 and 10 million steps they find more models than the transfer agents and continue to increase the number of found models exponentially, which becomes more linear to the end. After 15$\cdot 10^6$ the best performing experiments found about the same number of models as the best experiments run on 5256 for $50 \cdot 10^6$ steps. In average random exploration is outperformed by a factor of 700 for both manifolds. We note, that the decrease in the number of time steps appears to improve the efficiency of experiments run on manifolds with $h^{(1,1)} = 5$, as on neither manifold the experiments show the plateauing behaviour observed after $20 \cdot 10^6$  steps for manifold 5256.

	\section{Conclusion and Outlook}
	
	\label{sec:out}
	
	In this paper we constructed two environments representing heterotic line bundle models on CICYs for exploration with reinforcement learning algorithms. We showed that A3C agents outperform random walkers in identifying interesting vacuum configurations by factors up to 1700. Especially the flipping environment appears to be a favourable setting for the agents to develop long term strategies. This was also observed in the recent study of reinforcement learning of type IIA string vacua \cite{Halverson:2019tkf}, and we find that our results support their observation that reinforcement learning is able to pick up hidden patterns in the configurations space. Heterotic line bundle models on CICYs appear to be an even more successful setting for the application of A3C agents which here perform better by one order of magnitude compared to the IIA setting when taking random exploration as baseline.
	
	The results of Section \ref{sec:exp} show that deep reinforcement learning is able to reproduce many of the models found in the comprehensive scan \cite{Anderson:2013xka}. They further suggest that the agents scale well with larger $h^{(1,1)}$ (indeed, the run time for experiments vary very little between the $h^{(1,1)}=5$ and $h^{(1,1)}=6$ manifolds) and are thus particularly suited to study so far unexplored regions on manifolds that are computationally unfeasible to study systematically. We intend to come back to this question in the future.
	
	In Section \ref{sec:transfer} it has been demonstrated that the strategies learned by the agents are partly of general nature. The training obtained on one manifold allows an agent to also outperform random exploration on two other manifolds, one of which is topologically different from the first one. Additionally employing an agent with initialized weights according to a pretrained network and only adjusting the latter layers weights further improved the number of found models, but without reaching the performance of an agent fully trained on the new CICY.  We conclude that this is a positive indication of the agent's ability to identify general structures in string compactifications, thus clearly transcending human developed search strategies, but that experiments with different design are needed to corroborate this conclusion. A further motivation for such studies is that transfer learning should simplify future studies on larger $h^{(1,1)}$ on which the initial training will take more time. 
	
	Our study leads to several open questions and ideas for future studies. We remark that the use of transfer learning in this paper is slightly non standard. Most often, transfer learning is used in settings with convolutional neural networks rather than, as we do here, fully connected ones. In CNNs the first layers pick up some general features of the input space, which are then interpreted by fully connected layers later on. Given our positive indications on transfer learning, it would be interesting to apply CNNs in our environments. A natural choice for CNNs would be to use kernels scanning over the $h^{(1,1)}$ projective spaces or the five line bundles. Here one could potentially make use of an inception inspired architecture \cite{szegedy2014going}. Moreover, it has recently been observed that line bundle cohomologies on Calabi Yau $n$-folds are governed by degree $n$ polynomials \cite{Brodie:2019dfx,Brodie:2019ozt,Brodie:2019pnz,Larfors:2019sie,Klaewer:2018sfl,Constantin:2018hvl}. This further suggests the use of CNNs and an extension of the observation space to a matrix of shape $(5,h^{1,1},3)$ where the last dimension now contains the charges $(q_j^a)^d$ for $d \in {1,2,3}$.
	
	It would also be interesting to investigate the performance of other agents in our environments. A promising choice are Actor Critic using Kronecker-Factored Trust Region agents \cite{wu2017scalable} which are GPU trainable and supposedly more sample efficient than the A3C agents used in this paper. This is expected to greatly improve the analysis, as the agents have to explore through many configurations in order to find their first realistic model. This will also likely limit the number of time steps needed, which is very desirable, since the current computation of the reward is quite expensive.\footnote{The authors are currently working on an efficiency upgrade, which should partly remove this bottleneck.}
	
	In a final word we invite the interested reader to run their own experiments with new agents, different manifolds, different reward structures or just different hyperparameters. The code developed for this study is freely available and can be found online \cite{gymCICY}.
	
	\section*{Acknowledgements}
	
	RS would like to thank the organizers of the Deep Learning course at Uppsala University, in which this project originated. Both authors are grateful for valuable feedback and interesting discussions with F. Ruehle. The experiments were carried out on resources provided
	by the Swedish National Infrastructure for Computing (SNIC) at the Tetralith cluster of the National Supercomputer Centre (NSC) at Linköping University. This work is financed by
	the Swedish Research Council (VR) under grant numbers 2016-03873 and 2016-03503.

	\begin{appendix}

		\section{Heterotic $SU(5)$ GUT Models}\label{ap:a}
		
		In this appendix we provide a short review of the theory behind the physical constraints listed in section \ref{sec:rewards}. This review is included for completeness and intended for readers who are unfamiliar with this branch of string phenomenology.  A more thorough discussion of the framework of smooth heterotic GUT models can be found in \cite{Green:1987mn} and the heterotic line bundle papers \cite{Anderson:2011ns,Anderson:2012yf,Anderson:2013xka}.
		
		In this paper, we use RL to identify vector bundles that in compactifications of the heterotic string satisfy constraints needed for 4D, minimally supersymmetric, $SU(5)$ GUT models. These constraints are neither necessary nor sufficient, in the general case. However, for the setting we study, where the internal space of the compactification is a CY manifold and the vector bundle is a sum of line bundles, the constraints are necessary (but not sufficient). 
		
		Let us take a step back to 10D heterotic string theory to explain why we choose this setting. This theory has ${\cal N}=1$ supersymmetry in 10D, i.e. 16 supercharges. If we would compactify on a very simple space, such as a 6D torus, this would result in a theory with maximal supersymmetry. To obtain minimal 4D supersymmetry, we should instead compactify on a manifold that only allows one globally defined, covariantly constant spinor. CY manifolds satisfy this property, and are moreover Ricci-flat, thus solving Einstein's equations in vacuum. Consequently, we get a supersymmetric solution of heterotic string theory. CY manifolds have been constructed in great numbers using tools from algebraic geometry, and thus provide a plethora of possible compactifications.
		
		The particle physics sector of heterotic string compactifications is completely encoded in the vector bundle. To preserve supersymmetry, this vector bundle must admit a connection satisfying the Hermitian Yang--Mills equation. By the Donaldson--Uhlenbeck--Yau theorem, it is known that a holomorphic vector bundle $V$ over a CY manifold admits such a connection if  is polystable and have zero slope, $\mu(V)=0$. The simplest solution to the zero slope condition is that the bundle has vanishing first Chern class, which follows from condition \ref{c:sun} in our list.
		
		Polystability is then checked by showing that all coherent sub-sheaves ${\cal G} \subset V$ have inferior slope to $V$. These conditions are notoriously hard to check for generic bundles. For sums of line bundles $V = \bigoplus_{a} L_a$, on the other hand, the stability check simplifies (see \cite{Anderson:2012yf} for a further discussion). The slope of all constituent line bundles must vanish on a common locus in the K\"ahler moduli space of the CICY; this corresponds to  conditions \ref{c:weak}  and  \ref{c:kaehler} in our list.

		Finally, there is a 10D Bianchi identity, which act as a compatibility constraint for pairs $(X,V)$ of CY geometries $X$ and vector bundles $V$. The integrability constraint derived from this Bianchi identity puts limitations on the  second Chern class of the bundle, $c_2(V) \le c_2(X)$. For zero-slope, polystable bundles $V$ on $X$, the Bogomolov bound further constrains $c_2(V) > 0$. Both these constraints are listed in \ref{c:bianchi}.
		
		We now turn to $SU(5)$ GUT model specifics.  The 4D gauge group is specified by structure group of the vector bundle, which is $S(U(1)^5)$ for a sum of line bundles with vanishing first Chern class (condition \ref{c:sun} in our list). Starting from heterotic string with gauge group $E_8 \times E_8$, this results in a 4D gauge group which is $SU(5) \times U(1)^4$. We will not discuss these extra $U(1)$ factors in detail, just remark that they are often broken at high energy scales, and refer the reader to \cite{Anderson:2011ns,Anderson:2012yf,Anderson:2013xka} for more references on this issue. 
		
		We then want to ensure that the $SU(5)$ group can be broken to the Standard Model gauge group $SU(3)\times SU(2) \times U(1)$. This is possible, using Wilson lines, if the CICY has a freely acting discrete symmetry. We thus restrict to such CICYs; the manifolds studied in this paper all satisfy this constraint.
		
		Subsequently, the spectrum of the compactifcation, should, after Wilson line breaking, reduce to the three light fermion generations of Standard Model, without further exotic matter. The low-energy part of the spectrum is captured by massless modes of relevant 10D operators, cf. \cite{Green:1987mn}, and the number of such modes is in turn given by the dimension of bundle-valued cohomology groups (see \cite{Anderson:2013xka}). This leads to the last constraints in our list: three generations require \ref{c:index}, \ref{c:Index}, and \ref{c:fermion} (as least in the simple case where we disregard models that also include antigenerations). Finally, we require that there are no Higgs triplets, and the presence of at least one Higgs doublet, which is encoded in conditions \ref{c:triplet} and \ref{c:doublet}, respectively (see \cite{Anderson:2013xka} for more details).
		
	\end{appendix}

	\bibliography{ref}
	\bibliographystyle{ytphys}

	\newpage

	\appendix

\end{document}